\documentclass[twocolumn,english,showkeys,showpacs,superscriptaddress]{revtex4-1}
\usepackage[LGR,T1]{fontenc}
\usepackage[latin9]{inputenc}
\usepackage{amsmath}
\usepackage{amssymb}
\usepackage{graphicx}

\makeatletter

\ProvideTextCommand{\~}{LGR}[1]{\char126#1}

\usepackage[normalem]{ulem}

\usepackage{color}

\usepackage{ulem}

\newcommand{\stkout}[1]{\ifmmode\text{\sout{\ensuremath{#1}}}\else\sout{#1}\fi}

\makeatother

\usepackage{babel}
\usepackage{caption}
\begin{document}
\title{Superradiant Echoes Induced by Multiple Re-phasing of 
 NV Spin Sub-ensembles Grating at Room Temperature}

\author{Qilong Wu}
\address{Henan Key Laboratory of Diamond Optoelectronic Materials and Devices, Key Laboratory of Material Physics, Ministry of Education, School of Physics and Microelectronics, Zhengzhou University, Daxue Road 75, Zhengzhou 450052, China}

\author{Yuan Zhang}
\email{yzhuaudipc@zzu.edu.cn}
\address{Henan Key Laboratory of Diamond Optoelectronic Materials and Devices, Key Laboratory of Material Physics, Ministry of Education, School of Physics and Microelectronics, Zhengzhou University, Daxue Road 75, Zhengzhou 450052, China}
\address{Institute of Quantum Materials and Physics, Henan Academy of Sciences, Mingli Road 266-38, Zhengzhou 450046}

\author{Huihui Yu}
\address{Henan Key Laboratory of Diamond Optoelectronic Materials and Devices, Key Laboratory of Material Physics, Ministry of Education, School of Physics and Microelectronics, Zhengzhou University, Daxue Road 75, Zhengzhou 450052, China}

\author{Chong-Xin Shan}
\email{cxshan@zzu.edu.cn}
\address{Henan Key Laboratory of Diamond Optoelectronic Materials and Devices, Key Laboratory of Material Physics, Ministry of Education, School of Physics and Microelectronics, Zhengzhou University, Daxue Road 75, Zhengzhou 450052, China}
\address{Institute of Quantum Materials and Physics, Henan Academy of Sciences, Mingli Road 266-38, Zhengzhou 450046}

\author{Klaus M{\o}lmer}
\email{klaus.molmer@nbi.ku.dk}
\address{Niels Bohr Institute, University of Copenhagen, Blegdamsvej 17, 2100 Copenhagen, Denmark}

\begin{abstract}
In this Letter, we propose that superradiant echoes can be achieved at room temperature by applying a laser illumination and a microwave Hahn echo sequence to a diamond with a high concentration of nitrogen-vacancy (NV) centers placed in a dielectric microwave cavity. We identify that the combined action of two microwave driving pulses and a free evolution imprints a phase grating among NV spin sub-ensembles in frequency space, and multiple re-phasing of the grated spin sub-ensembles leads to multiple superradiant echoes through a collective coupling with the cavity.  Furthermore, we show that the superradiant echoes can be actively tailored with the microwave pulses and the laser illumination by modifying the grating parameters, and the multiple re-phasing dynamics is analogous to the one leading to superradiant beats in optical clock system. In the future, the spin sub-ensembles grating and the resulting echoes can be further optimized with dynamical decoupling, which might pave the way for applications in quantum sensing. 
\end{abstract}
\maketitle

\emph{Introduction---} Electron paramagnetic resonance (EPR) is a technique for detecting unpaired electrons with applications in chemistry ~\citep{Roessler,Fan}, medicine ~\citep{Misra}, and quantum sensing ~\citep{Qin,Simpson,Staudacher}. Pulsed EPR applies microwave pulses to manipulate spin states of the electrons and then detect the generated echoes ~\citep{Schweiger}, among which microwave cavities are often employed to enhance the coupling of the sample with the microwave field. In particular, superconducting resonators support microwave modes with high quality factors, and bring the electron spins-cavity system into strong coupling regime, where Rabi splitting~\citep{Putz,Angerer} and Rabi oscillations~\citep{Putz,Krimer} appear due to the formation of spins-photons hybrid modes and coherent spins-photons energy exchange, respectively.

Although continuous-wave EPR has been studied extensively in the strong coupling regime ~\citep{Wallraff,Khitrova}, pulsed EPR is less explored so far. In 2017, Putz {\it et al.} observed several echoes when applying a standard Hahn echo sequence to nitrogen vacancy (NV) center spins in a diamond strongly coupled with a superconducting resonator at cryogenic temperature~\citep{SPutz}. Later, in 2020, Weichselbaumer {\it et al.}~\citep{WeichselbaumerS} and Debnath {\it et al.} ~\citep{DebnathK} reported same effects using phosphorus spins and neodymium ions. Recently, de Graaf {\it et al.}~\citep{Graaf} studied the dependence of this effect on the driving pulse parameters systematically. Debnath {\it et al.} interpreted the effect as caused by back-action of echoes on individual spins, and refocusing of individual spins at later times, which is termed as self-stimulated echoes (SSEs). Although such an interpretation delivers the key ingredient of the mechanism, further studies are required to reveal the excitation and synchronization of spins in different frequency classes, and the role of the collective coupling with the cavity. Furthermore, all the experiments were carried out so far at cryogenic temperature, which begs the question whether they can also be observed at room temperature.

In parallel with the research on SSEs, there were also extensive studies of superradiant lasing ~\citep{Bohnet,Norcia2}. In contrast to conventional lasers relying on optical coherence and good cavities, superradiant lasers depend on the coherence among excited atoms and bad cavities. Since the superradiant laser has extremely narrow linewidth and is robust against cavity fluctuations, it has potential applications in optical clocks and quantum metrology~\citep{ADLudlow}. Along this line, Norcia {\it et al.} reported superradiant pulses from strontium optical clock transition in 2016~\citep{Norcia1} and active frequency measurement based on these signals in 2018~\citep{Norcia3}. Meanwhile, they also observed superradiant beats from two and ten atomic sub-ensembles~\citep{Norcia3}. Interestingly, we observe a strong similarity between the superradiant beats and the SSEs, see Fig. S1 of the Supplemental Material (SM).

\begin{figure}[!htp]
\includegraphics[scale=1]{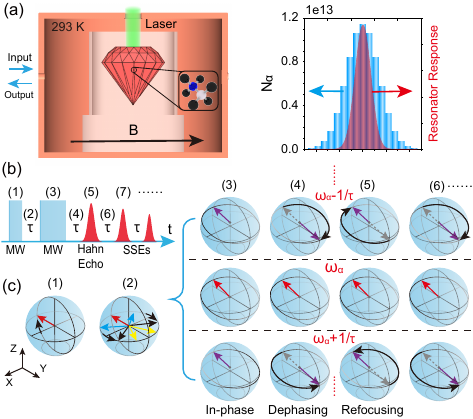}
\caption{\label{fig:system} System schematics and superradiant echoes. (a) illustrates a bulk diamond inside a dielectric ring surrounded by a copper cavity in a strong static magnetic field at room temperature (left), where the NV centers (inset) are excited by a laser, the microwave field drives the system and the reflected field is detected. The right part shows the relationship between the inhomogeneous profile of the NV spin transition frequencies and the cavity response. (b) shows the Hahn echo sequence, where two microwave driving pulses (blue) and superradiant echoes (red) are separated by a free evolution time $\tau$. (c) illustrates the Bloch vector dynamics of spin sub-ensembles: the microwave driving pulses and the free evolution result in groups of spin sub-ensembles with frequency spacing $1/\tau$ (forming a grating), the initially in-phase spin sub-ensembles across groups become out-of-phase and then in-phase at time $n\times\tau$.} 
\end{figure}

To provide further insights into the SSEs, avoid the complexity of cryogenic experiments and reveal the analogy with the superradiant beats, in this Letter, we propose that superradiant echoes can be achieved at room temperature by shining a laser on a bulk diamond with a high concentration of NV centers inside a dielectric microwave cavity [Fig.~\ref{fig:system}(a)] and applying a standard Hahn echo sequence [Fig.~\ref{fig:system}(b)]. In contrast to Debnath's interpretation with individual NV spins in the time domain ~\citep{DebnathK}, we divide the NV spins into multiple sub-ensembles with different transition frequencies to account for the inhomogeneous broadening [right part of Fig.~\ref{fig:system}(a)], and employ Dicke states and collective Bloch vectors to analyze their dynamics. We achieve a unified picture of the spin sub-ensembles dynamics leading to all the echoes, as schematically shown in Fig.~\ref{fig:system}(c). We identify that the combined effect of two microwave driving pulses and a free evolution of length $\tau$ imprints a phase grating among spin sub-ensembles in frequency domain, leading to multiple groups of sub-ensembles with frequency spacing $f=1/\tau$ and in-phase condition for corresponding sub-ensembles across groups. Although these sub-ensembles become out-of-phase later on, they can become in-phase again at times $n\times\tau$, and then their collective coupling with the cavity results in the superradiant echoes. The latter dynamics bears great similarities with that leading to the superradiant beats of optical clock atoms. Furthermore, the laser power and the microwave pulses can be used to actively control the superradiant echoes by modifying the spin sub-ensembles grating.

\emph{System and Theory---} We consider a bulk diamond inside a dielectric cavity, in which an ensemble of NV center spins couples strongly with a microwave mode [Fig.~\ref{fig:system}(a)]. At room temperature, the spin states reach thermal equilibrium due to strong spin-lattice relaxation through two-phonon Raman scattering and the Orbach process~\citep{JarmolaA}. By shining laser on the diamond, optically induced spin polarization can compensate the spin-lattice relaxation, and cool the NV spins to states equivalent to those at low temperature. The NV spins are hence prepared near Dicke ground states with high symmetry~\citep{Yuan1,Yuan2,HWang}, and the enhanced coupling with the cavity establishes the basis for observing the superradiant echoes.

We adopt first-order mean-field equations to describe the system dynamics (see Sec. S2 of the SM), and use the parameters from a recent experiment~\citep{TomDay}. The microwave cavity is modeled as a harmonic oscillator with a frequency $\omega_c = 2\pi \times 9.8$ GHz and a photon damping rate $\kappa = \kappa_1 + \kappa_2$, which includes a coupling loss $\kappa_1 = 2\pi \times 0.95$ MHz and an internal loss $\kappa_2 = 2\pi \times 0.89$ MHz. The cavity is driven by a microwave field with frequency $\omega_d \approx \omega_c$ and a power of $12$ dBm, which leads to a driving rate $\Omega = 2\pi \times8\times10^9$ Hz$^{-1/2}$. For simplicity, we consider only the NV spin levels that couple resonantly with the microwave cavity, and treat them as two-level systems~\citep{Yuan4}. To account for the inhomogeneous broadening, we divide the whole ensemble (about $7.3 \times 10^{13}$ spins) into $N=2000$ sub-ensembles (indexed by $\alpha$), and assume that the number of spins $N_\alpha$ follows a Gaussian distribution $dN_\alpha/d\omega_\alpha = [N/\sqrt{2\pi\sigma^2}]e^{-(\omega_\alpha -\omega_s)^2/(2\sigma^2)}$ with the spin transition frequency $\omega_\alpha$ [right of Fig.~\ref{fig:system}(a)]. The distribution centers around the frequency of the microwave cavity ($\omega_s=\omega_c$), and has a linewidth of $\Gamma =  2\sqrt{\rm 2 ln2}\sigma = 2\pi \times 3.3$ MHz. For simplicity, we discretize uniformly the whole ensemble by setting  $\omega_{\alpha}=\omega_s + (\alpha -N/2)\Delta$ with the frequency spacing $\Delta=0.04$ MHz between the neighboring sub-ensembles. We assume that the spins in each sub-ensemble couple with the cavity with the same strength $g_\alpha=2\pi\times0.18$ Hz, and express also the relaxation, the optically-induced polarization and the dephasing with the rates $\gamma_\alpha = 2\pi\times23.7$ Hz, $\eta_\alpha=0.1$ Hz $\sim 0.1$ MHz, $\chi_\alpha = 2\pi\times0.014$ MHz, respectively.

To understand the dynamics leading to the superradiant echoes, we consider the collective Bloch vector ${\bf J}_\alpha = \sum_{i=x,y,z} J_{i_\alpha} {\bf e}_i$ with the components $J_{x_\alpha}$,$J_{y_\alpha}$, and $J_{z_\alpha}$ and the unit vectors of the Cartesian coordinate system ${\bf e}_i$. We also introduce the Dicke states $\left|J_\alpha,M_\alpha\right\rangle$ for each sub-ensemble, and we associate the mean Dicke quantum numbers $\bar{J}_\alpha, \bar{M}_\alpha$ with the collective Bloch vector through the relations $\bar{J}_\alpha \approx \sqrt{\sum_{i=x,y,z} \langle \hat{J}_{i_\alpha}\rangle^{2}}, \bar{M}_\alpha  = \langle \hat{J}_{z_\alpha}\rangle$. The Dicke states have integer or half-integer quantum numbers $J_\alpha$ in the range $0(1/2)\le J_\alpha \le N_\alpha/2$, $-J_\alpha \le M_\alpha \le J_\alpha$, where $J_\alpha$ characterizes the collective coupling strength with the cavity mode, and the number $M_\alpha$ indicate the degree of excitation of the sub-ensembles~\citep{Yuan1}.

\begin{figure}[ht!]
\begin{centering}   
\includegraphics[scale=1]{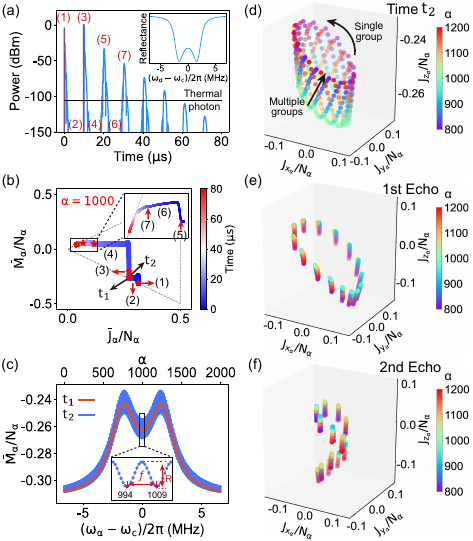}
\par\end{centering}
\caption{\label{fig:dynamics} Room-temperature superradiant echoes and spin sub-ensembles dynamics. (a) shows a train of echoes (blue peaks) induced by two microwave driving pulses (red spikes), which become gradually weaker than  the thermal photon noise power (horizontal line). The numbers (1-7) refer to the time points marked in Fig.~\ref{fig:system}(b). The inset shows the normalized reflectance as a function of the frequency detuning of the microwave driving to the cavity. (b) presents the dynamics within the Dicke state space for the spin sub-ensemble perfectly resonant with the cavity. $t_1$ and $t_2$ indicate the start and midpoint of the second driving pulse.  (c) depicts the average excitation number of Dicke states $\bar{M}_\alpha/N_\alpha$ as a function of the frequency detuning of the spin sub-ensembles to the microwave cavity (lower axis) and the ensemble index $\alpha$ (upper axis) at time $t_1$ and $t_2$ (red and blue lines). The inset shows the zoomed area, where the excitation grating can be characterized by the frequency span $f=1/\tau$ and the amplitude $R$. (d)-(f) show the end-points of the Bloch vectors for several groups of spin sub-ensembles, which are near-resonant or resonant with the cavity, at time $t_2$ and the times when the first and second echo occurs, respectively. Here, the Bloch vector components $J_{x_\alpha},J_{y_\alpha},J_{z_\alpha}$ are normalized by the number of spins in each sub-ensemble $N_\alpha$. In the simulation, the optically induced cooling rate $\eta_{\alpha} =5 \times 10^2$ Hz.}
\end{figure}

\emph{Mechanism of Superradiant Echoes---} In Fig.~\ref{fig:dynamics}, we explore the system dynamics at room temperature under continuous laser illumination. The reflection spectrum for a weak continuous-wave microwave probe shows two dips separated by about $2\pi\times 3$ MHz [inset of Fig. \ref{fig:dynamics}(a)], confirming the strong coupling condition for the current system.  The emitted microwave field after excitation by two microwave pulses, which are $28$ ns and $56$ ns long respectively, and separated by $\tau$ = $10$ $\mu$s, shows several echoes with a period of $\tau$ after the first two peaks [Fig. \ref{fig:dynamics}(a)], which suggests the feasibility of observing superradiant echoes at room temperature. Here, we assume that the echoes above the power of thermal photons (horizontal line) can be observed in the experiment. 

To understand the mechanism underlying the superradiant echoes, we analyze the dynamics of spin sub-ensembles with both the Dicke states and the Bloch vectors. As an example, we show the dynamics of the spin sub-ensemble which is perfectly resonant with the microwave cavity [Fig. \ref{fig:dynamics}(b)]. Due to the optically induced spin polarization, this sub-ensemble is initially prepared to a mixed state near the middle of lower boundary of Dicke state space. It is driven vertically by the first driving pulse, and then evolves almost horizontally towards the lower boundary due to dephasing. After that, the spin sub-ensemble moves vertically again towards states with positive mean Dicke quantum number $\bar{M}_\alpha>0$ by the second driving pulse, and it finally evolves horizontally towards the upper boundary of the Dicke state space. Other spin sub-ensembles follow similar dynamics except that some reach the states with $\bar{M}_\alpha<0$ after the second driving pulse [see Sec. S4.1 of the SM]. 

By plotting $\bar{M}_\alpha$ (normalized by the number of spins $N_\alpha$) as a function of the frequency detuning of the spin sub-ensembles to the microwave cavity, we obtain the results shown in Fig.~\ref{fig:dynamics}(c) at the end of the first free evolution (red curve) and during the second driving pulse (blue curve). In the former case, we find two symmetric peaks, which resemble that of the reflection spectrum shown in Fig.~\ref{fig:dynamics}(a) and can be attributed to the strong coupling. In contrast, in the latter case, the excitation in the system depends on the detuning in a rapidly oscillating manner, see Sec. S4.2 of the SM for more details. Inspired by this grating, we can combine the spin sub-ensembles within a single oscillation to form a single group, and split the whole spin ensemble into multiple groups.  Furthermore, we characterize the group in the middle of inhomogeneous profile with the oscillation amplitude $R$ and frequency span $f$ where the latter is in excellent agreement with the inverse of the free evolution time $f=1/\tau$, as shown in Sec. S4.3 of the SM.

The Bloch vectors of multiple spin sub-ensembles behave as follows. By applying the first driving pulse, the Bloch vectors rotate around the x-axis to a plane below the equatorial plane. During the free evolution, the Bloch vectors spread out due to different transition frequencies of spin sub-ensembles, and also become shortened due to the dephasing. When applying the second driving pulse, the Bloch vectors are elevated and rotated slightly around the x-axis. As a result, the end-points of the vectors form a circle for a single group of spin sub-ensembles, and the circles for different groups form a spiral [Fig.~\ref{fig:dynamics}(d)]. Since the projections of the Bloch vectors in the equatorial plane are more or less aligned for some sub-ensembles across the groups, these sub-ensembles are in-phase, and the oscillation pattern as revealed above reflects a phase grating in frequency space among the spin sub-ensembles. Such a phase grating can be also perceived by analyzing the Bloch vector components $J_{x_\alpha},J_{y_\alpha}$ as functions of frequency detuning and time (see Sec. S4.4 of the SM).  

After the second driving pulse, the Bloch vectors of the spin sub-ensembles become spread and shorted again due to the different frequencies and the dephasing. However, at the delay time $\tau$, the Bloch vectors become refocused for the spin sub-ensembles across groups, which are separated by $m\times f$ ($m$ is an integer) in frequency space, and the vectors are above and below the horizontal plane for half of the sub-ensembles [Fig.~\ref{fig:dynamics}(e)]. When the spin sub-ensembles interact the cavity, the upper  and lower half of spin sub-ensembles can emit and absorb collectively the microwave field, respectively. However, since there is almost no microwave field initially, the emission dominates over the absorption, leading to the first echo. Then, the above dynamics repeats in time, and the Bloch vectors become refocused again at the delay time $2\tau$ [Fig.~\ref{fig:dynamics}(f)], and the second echo is generated. Since the spin sub-ensembles interact quite strongly with the cavity mode, the decayed field of the former echo acts also back on the spin sub-ensembles, and causes the binding of single circle to two circles, and further to three circles. By combining the above analysis, we derive a unified picture of the dynamics for the self-stimulated echoes or the superradiant echoes, as shown in Fig.~\ref{fig:system}(c). Furthermore, we have also simulated superradiant echoes in a system in a deep strong coupling regime at room temperature~\citep{SPutz}, where the spin sub-ensembles grating is clearly observed but the Bloch vectors dynamics becomes a mess, see Sec. S4.5 of the SM.

\begin{figure}[ht!]
\begin{centering}   
\includegraphics[scale=1]{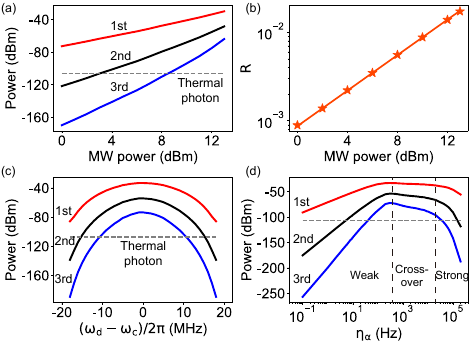}
\par\end{centering}
\caption{\label{fig:sse} Tailoring of superradiant echoes.  (a,c,d) show the variation of the output power of the first three echoes with the microwave pulses power (a) and frequency (c), as well as the optically induced spin polarization rate $\eta_\alpha$ (d). (b) illustrates the corresponding changes of the grating amplitude $R$, as marked in Fig.~\ref{fig:dynamics}(c). For panels (a) and (c), we assume $\eta_\alpha=5\times10^2$ Hz.}
\end{figure}

\emph{Tailoring of Superradiant Echoes---} In the following, we demonstrate that both the number and amplitude of echoes can be actually tailored by controlling the properties of the grating. Figure~\ref{fig:sse} (a,b) show that as the microwave power increases from $0$ to $12$ dBm, the grating amplitude increases, resulting in the increase of the echo power. However, considering the thermal noise, the second and third echo can only be observed experimentally when the microwave driving exceeds $3$ dBm and $9$ dBm, respectively.  Figure~\ref{fig:sse}(c) indicates that by increasing the frequency detuning of the microwave driving field to the cavity from $-2\pi\times20$ MHz to $2\pi\times20$ MHz, the power of the echoes follows a parabolic curve with a maximum under the resonant condition, and the second and third echo are only visible for smaller range of frequency detuning. Figure.~\ref{fig:sse}(d) presents that as the optically induced polarization rate $\eta_\alpha$ increases from $0.1$ Hz to $0.1$ MHz, the amplitude of echoes first increases for $\eta_\alpha<400$ Hz, becomes saturated within the range $400<\eta_\alpha<2\times10^4$ Hz, and starts decreasing for $\eta_\alpha>2\times10^4$ Hz, where the linear, saturation, and suppression of echoes are related to the system in the weak, crossover and strong coupling regime (see Sec. S4.6 of the SM). Under the strong coupling, the microwave excitation is suppressed due to the photon blockade effect~\citep{Boit}, and thus the amplitude of the echoes becomes reduced.

\begin{figure}[ht!]
\begin{centering}   
\includegraphics[scale=1]{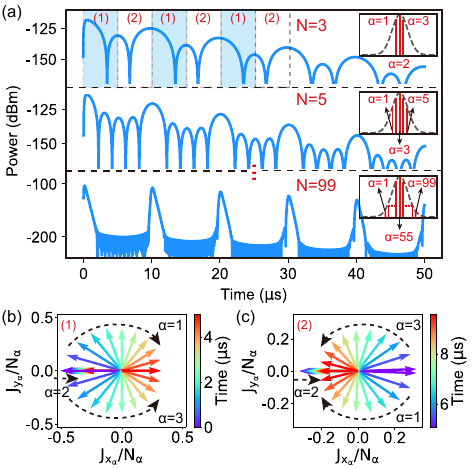}
\par\end{centering}
\caption{\label{fig:superbeats} Analogy of superradiant echoes to superradiant beats. (a) shows the superradiant beats for a simplified system with 3, 5, and 99 discrete spin sub-ensembles with same frequency spacing $f=1/\tau$ under the spin inhomogeneous profile (insets). (b) and (c) show the evolution of the Bloch vectors for the three spin sub-ensembles at one period as marked on the top of the panel (a).} 
\end{figure}

\emph{Analogy to Superradiant Beats---} After understanding the mechanism behind the superradiant echoes, we now demonstrate that part of the mechanism is analogous to the one leading to the superradiant beats in the optical clock system. To this end, we concentrate on those spin sub-ensembles located at the peaks of the spin sub-ensembles grating, and mimic them with discrete sub-ensembles with equal frequency spacing $f=1/\tau$ [insets of Fig.~\ref{fig:superbeats}(a)], which are prepared in a pure state such that the Bloch vectors point along the negative x-axis, i.e. the initial in-phase among the sub-ensembles. For the simplest system with three sub-ensembles, in a frame rotating with the frequency of the middle sub-ensemble, the collective Bloch vector is fixed along the negative x-axis for this ensemble, and that of the other two sub-ensembles rotate clockwise and anti-clockwise in the equatorial plane with same period $1/f$ [Fig.~\ref{fig:superbeats}(b,c)]. When the Bloch vectors become refocused for two or three sub-ensembles, the spins couple with the microwave cavity and lead to weak and strong superradiant peaks, as marked on the top of Fig.~\ref{fig:superbeats}(a). When this refocusing dynamics repeats in time, the radius of the Bloch vectors decreases due to the dephasing, resulting in decreased peaks.

If more spin sub-ensembles are added to the system, their Bloch vectors will rotate faster at a double frequency $2f$, a triple frequency $3f$, or multiple times of the fundamental frequency $nf$ (with integer $n$). Although the refocusing of these Bloch vectors becomes more complex, we can easily deduce that they can be completely refocused only at the time $n \tau = n/f$. Then, the superradiant beats become stronger and narrower [middle and bottom of Fig.~\ref{fig:superbeats}(a)], leading to the similar feature as the superradiant echoes. These results suggest that the Bloch vectors refocusing as the kernel of the superradiant echoes is also responsible for the superradiant beats.  

\emph{Conclusions---} In summary, our study suggests that superradiant echoes can be achieved at room temperature by applying a Hahn echo sequence to a strongly coupled system with NV center spins and a dielectric microwave cavity under laser illumination. Our calculation shows that the system evolution due to two microwave pulses and a free evolution in between results in a phase grating with period $f=1/\tau$ among spin sub-ensembles in frequency space, and the sub-ensembles separated by $m\times f$ in frequency become re-phased at time $n\tau$ and thus couple strongly with the cavity, leading to multiple superradiant echoes. In addition, the microwave pulses and the laser power can be tailored to actively control the superradiant echoes by modifying the amplitude of the phase grating, and the re-phasing dynamics of spin sub-ensembles is analogous to the mechanism leading to the superradiant beats of optical clock atoms.  

Our study not only explains the mechanism behind the previous experimental observations, but also suggests that further exploration under ambient conditions might be feasible with solid-state spin systems, such as pentacene molecular spins \citep{wuhao}, silicon-vacancy centers in silicon carbide \citep{Widmann}, and boron vacancy centers in hexagonal boron nitride \citep{Liu}. In the future, a similar analysis can be applied to study the superradiant echoes under a dynamic decoupling sequence, which might pave the way for the application in quantum sensing.

\vspace{\baselineskip} 
\begin{acknowledgments}
Qilong Wu carried out the numerical calculations under the supervision of Yuan Zhang. They contribute equally to the work. All authors contributed to the analysis and writing of the manuscript. This work was supported by the National Key R\&D Program of China under grant 2024YFE0105200, the National Natural Science Foundation of China through the project No. 12422413, 62475242, the Cross-disciplinary Innovative Research Group Project of Henan Province No. 232300421004, as well as by the Carlsberg Foundation through the ``Semper Ardens'' Research Project QCooL. 
\end{acknowledgments}


\begin{thebibliography}{10}

\bibitem{Roessler} M. M. Roessler and E. Salvadori, Principles and applications of EPR spectroscopy in the chemical sciences, Chem. Soc. Rev. \textbf{47}, 2534 (2018).

\bibitem{Fan} T. W.-M. Fan and A. N. Lane, Applications of NMR spectroscopy to systems biochemistry, Prog. Nucl. Magn. Reson. Spectrosc. \textbf{18}, 92 (2016).


\bibitem{Misra} S. K. Misra, Multifrequency Electron Paramagnetic Resonance: Theory and Applications, 1st ed.,  Wiley VCH, New York, (2011).

\bibitem{Qin} Z. Qin, Z. Wang, F. Kong, J. Su, Z. Huang, P. Zhao, S. Chen, Q. Zhang, F. Shi, and J. Du, In situ electron paramagnetic resonance spectroscopy using single nanodiamond sensors, Nat. Commun. \textbf{14}, 6278 (2023).

\bibitem{Simpson} D. A. Simpson, R. G. Ryan, L. T. Hall, E. Panchenko, S. C. Drew, S. Petrou, P. S. Donnelly, P. Mulvaney, and L. C. L. Hollenberg, Electron paramagnetic resonance microscopy using spins in diamond under ambient conditions, Nat. Commun. \textbf{8}, 458 (2017).

\bibitem{Staudacher}  T. Staudacher, N. Raatz, S. Pezzagna, J. Meijer, F. Reinhard, C. A. Meriles, and J. Wrachtrup, Probing molecular dynamics at the nanoscale via an individual paramagnetic centre, Nat. Commun. \textbf{6}, 8527 (2015).

\bibitem{Schweiger} A. Schweiger and G. Jeschke, Principles of pulse electron paramagnetic resonance, Oxford University Press, New York (2001).

\bibitem{Putz} S. Putz, D. O. Krimer, R. Ams{\"u}ss, A. Valookaran, T. N{\"o}bauer, J. Schmiedmayer, S. Rotter, and J. Majer, Protecting a spin ensemble against decoherence in the strong coupling regime of cavity QED, Nat. Phys. \textbf{10}, 720 (2014).


\bibitem{Angerer} A. Angerer, T. Astner, D. Wirtitsch, H. Sumiya, S. Onoda, J. Isoya, S. Putz, and J. Majer, Collective strong coupling with homogeneous Rabi frequencies using a 3D lumped element microwave resonator, Appl. Phys. Lett. \textbf{109}, 033508 (2016).

\bibitem{Krimer} D. O. Krimer, S. Putz, J. Majer, and S. Rotter, Non-Markovian dynamics of a single-mode cavity strongly coupled to an inhomogeneously broadened spin ensemble, Phys. Rev. A \textbf{90}, 043852 (2014).

\bibitem{Wallraff} A. Wallraff, D. I. Schuster, A. Blais, L. Frunzio, R.-S. Huang, J. Majer, S. Kumar, S. M. Girvin, and R. J. Schoelkopf, Strong coupling of a single photon to a superconducting qubit using circuit quantum electrodynamics, Nature \textbf{431}, 162 (2004).

\bibitem{Khitrova} G. Khitrova, H. M. Gibbs, M. Kira, S. W. Koch, and A. Scherer, Vacuum Rabi splitting in semiconductors, Nat. Phys. \textbf{2}, 81 (2006).

\bibitem{SPutz} S. Putz, Circuit Cavity QED with Macroscopic Solid-State Spin Ensembles, PhD thesis, Springer, (2017).

\bibitem{WeichselbaumerS} S. Weichselbaumer, M. Zens, C. W. Zollitsch, M. S. Brandt, S. Rotter, R. Gross, and H. Huebl, Echo Trains in Pulsed Electron Spin Resonance of a Strongly Coupled Spin Ensemble, Phys. Rev. Lett. \textbf{125}, 137701 (2020).

\bibitem{DebnathK} K. Debnath, G. Dold, J. J. L. Morton, and K. M{\o}lmer, Self-Stimulated Pulse Echo Trains from Inhomogeneously Broadened Spin Ensembles, Phys. Rev. Lett. \textbf{125}, 137702 (2020).

\bibitem{Graaf} S. E. de Graaf, A. Jayaraman, S. E. Kubatkin, A. V. Danilov, and V. Ranjan, Scaling of self-stimulated spin echoes, Appl. Phys. Lett. \textbf{124}, 024001 (2024).

\bibitem{Bohnet} J. G. Bohnet, Z. Chen, J. M. Weiner, D. Meiser, M. J. Holland, and J. K. Thompson, A steady-state superradiant laser with less than one intracavity photon, Nature \textbf{484}, 78 (2012).


\bibitem{Norcia2} M. A. Norcia, M. N. Winchester, J. R. K. Cline, and J. K. Thompson, Superradiance on the Millihertz Linewidth Strontium Clock Transition, Sci. Adv. \textbf{2}, e1601231 (2016).


\bibitem{ADLudlow} A. D. Ludlow, M. M. Boyd, J. Ye, E. Peik, and P. O. Schmidt, Optical atomic clocks. Rev. Mod. Phys. \textbf{87}, 637 (2015). 

\bibitem{Norcia1} M. A. Norcia and J. K. Thompson, Cold-Strontium Laser in the Superradiant Crossover Regime, Phys. Rev. X \textbf{6}, 011025 (2016).


\bibitem{Norcia3} M. A. Norcia, J. R. K. Cline, J. A. Muniz, J. M. Robinson, R. B. Hutson, A. Goban, G. E. Marti, J. Ye, and J. K. Thompson, Frequency Measurements of Superradiance from the Strontium Clock Transition, Phys. Rev. X \textbf{8}, 021036 (2018).

\bibitem{JarmolaA} A. Jarmola, V. M. Acosta, K. Jensen, S. Chemerisov, and D. Budker, Temperature- and Magnetic-Field-Dependent Longitudinal Spin Relaxation in Nitrogen-Vacancy Ensembles in Diamond, Phys. Rev. Lett. \textbf{108}, 197601 (2012).

\bibitem{Yuan1} Y. Zhang, Q. Wu, S.-L. Su, Q. Lou, C. Shan, and K. M{\o}lmer, Cavity Quantum Electrodynamics Effects with Nitrogen Vacancy Center Spins Coupled to Room Temperature Microwave Resonators, Phys. Rev. Lett. \textbf{128}, 253601 (2022).

\bibitem{Yuan2} Y. Zhang, Q. Wu, H. Wu, X. Yang, S.-L. Su, C. Shan, and K. M{\o}lmer, Microwave mode cooling and cavity quantum electrodynamics effects at room temperature with optically cooled nitrogen-vacancy center spins, npj Quantum Inf. \textbf{8}, 125 (2022).

\bibitem{HWang} H. Wang, K. L. Tiwari, K. Jacobs, M. Judy, X. Zhang, D. R. Englund, and M. E. Trusheim, A spin-refrigerated cavity quantum electrodynamic sensor, Nat. Commun. \textbf{15}, 10320 (2024).


\bibitem{TomDay} T. Day, M. Isarov, W. J. Pappas, B. C. Johnson, H. Abe, T. Ohshima, D. R. McCamey, A. Laucht, and J. J. Pla, Fate of photon blockade in the deep strong-coupling regime, Phys. Rev. X \textbf{14}, 041066 (2024).


\bibitem{Yuan4} Q. Wu, Y. Zhang, X. Yang, S.-L. Su, C. Shan, and K. M{\o}lmer, A superradiant maser with nitrogen-vacancy center spins. Sci. China-Phys. Mech. Astron. \textbf{65}, 217311 (2022).

\bibitem{Boit}  A. L. Boit\'e, M. J. Hwang, H. Nha, and M. B. Plenio, Fate of photon blockade in the deep strong-coupling regime, Phys. Rev. A \textbf{94}, 033827 (2016).

\bibitem{wuhao} H. Wu, S. Mirkhanov, W. Ng, and M. Oxborrow, Bench-Top Cooling of a Microwave Mode Using an Optically Pumped Spin Refrigerator, Phys. Rev. Lett. \textbf{127}, 053604 (2021).


\bibitem{Widmann} M. Widmann, S. Y. Lee, T. Rendler, N. T. Son, H. Fedder, S. Paik, L. P. Yang, N. Zhao, S. Yang, I. Booker, A. Denisenko, M. Jamali, S. A. Momenzadeh, I. Gerhardt, T. Ohshima, A. Gali, E. Janz\'{e}n, and J. Wrachtrup, Coherent control of single spins in silicon carbide at room temperature, Nat. Mater. \textbf{14}, 164 (2015).


\bibitem{Liu} W. Liu, V. Iv\'{a}dy, Z.-P. Li, Y.-Z. Yang, S. Yu, Y. Meng, Z.-A. Wang, N.-J. Guo, F.-F. Yan, Q. Li, J.-F. Wang, J.-S. Xu, X. Liu, Z.-Q. Zhou, Y. Dong, X.-D. Chen, F.-W. Sun, Y.-T. Wang, J.-S. Tang, A. Gali, C.-F. Li, and G.-C. Guo, Coherent dynamics of multi-spin $V_B^-$ center in hexagonal boron nitride, Nat. Commun. \textbf{13}, 5713 (2022).
 


\end{thebibliography}
\end{document}


\title{Supplemential Material for \\ Superradiant Echoes Induced by Multiple Re-phasing of NV Spin Sub-ensembles Grating at Room Temperature}

\author{Qilong Wu}
\address{Henan Key Laboratory of Diamond Optoelectronic Materials and Devices, Key Laboratory of Material Physics, Ministry of Education, School of Physics and Microelectronics, Zhengzhou University, Daxue Road 75, Zhengzhou 450052, China}

\author{Yuan Zhang}
\email{yzhuaudipc@zzu.edu.cn}
\address{Henan Key Laboratory of Diamond Optoelectronic Materials and Devices, Key Laboratory of Material Physics, Ministry of Education, School of Physics and Microelectronics, Zhengzhou University, Daxue Road 75, Zhengzhou 450052, China}
\address{Institute of Quantum Materials and Physics, Henan Academy of Sciences, Mingli Road 266-38, Zhengzhou 450046}

\author{Huihui Yu}
\address{Henan Key Laboratory of Diamond Optoelectronic Materials and Devices, Key Laboratory of Material Physics, Ministry of Education, School of Physics and Microelectronics, Zhengzhou University, Daxue Road 75, Zhengzhou 450052, China}

\author{Chong-Xin Shan}
\email{cxshan@zzu.edu.cn}
\address{Henan Key Laboratory of Diamond Optoelectronic Materials and Devices, Key Laboratory of Material Physics, Ministry of Education, School of Physics and Microelectronics, Zhengzhou University, Daxue Road 75, Zhengzhou 450052, China}
\address{Institute of Quantum Materials and Physics, Henan Academy of Sciences, Mingli Road 266-38, Zhengzhou 450046}

\author{Klaus M{\o}lmer}
\email{klaus.molmer@nbi.ku.dk}
\address{Niels Bohr Institute, University of Copenhagen, Blegdamsvej 17, 2100 Copenhagen, Denmark}

\maketitle
\tableofcontents

\newpage

\renewcommand\thefigure{S\arabic{figure}}
\renewcommand\thetable{S\arabic{table}}
\renewcommand{\thepage}{S\arabic{page}}
\renewcommand{\bibnumfmt}[1]{[#1]}
\setcounter{figure}{0}  
\renewcommand{\theequation}{S\arabic{equation}}
\renewcommand{\thesection}{S\arabic{section}}
\renewcommand{\thetable}{S\arabic{table}}
\renewcommand{\thesubsection}{\arabic{subsection}}  
\section{Similarity Between Self-stimulated Echoes and Superradiant Beats \label{sec:comp}}

\begin{figure}[ht!]
\begin{centering}   
\includegraphics[scale=1]{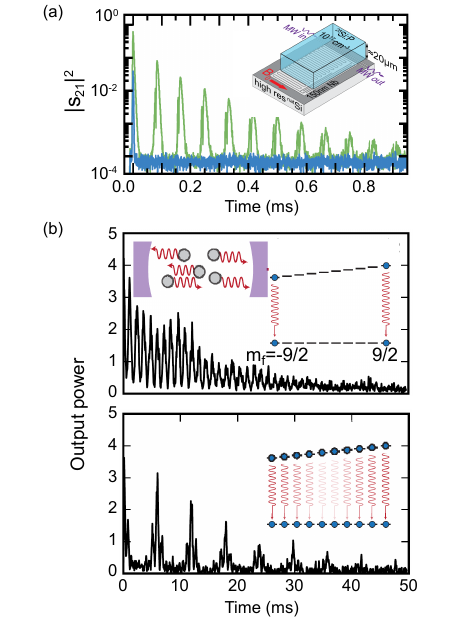}
\par\end{centering}
\caption{\label{fig:echoes-beats} Comparison of self-stimulated echoes of phosphorus spin ensembles coupled to a superconducting microwave resonator (a), and superradiant beats of strontium atoms coupled to an optical cavity (b), which were studied experimentally in Refs.~\citep{WeichselbaumerS,Norcia3}, respectively. In panel (b), the upper and low parts show the results for the system with two and ten atomic sub-ensembles. The figures are adapted from Refs.~\citep{WeichselbaumerS,Norcia3}. Reprinted with permission from Weichselbaumer {\it et al.} Phys. Rev. Lett. \textbf{125}, 137701 (2020). Copyright (2024) by the American Physical Society.}
\end{figure}

In the main text, we point out that the self-stimulated echoes, as recently observed in several experiments~\citep{WeichselbaumerS,DebnathK}, share many similarities with the superradiant beats of the optical clock transitions of strontium atoms~\citep{Norcia3}. To illustrate these similarities, we have plotted two figures from Refs.~\citep{WeichselbaumerS} and~\citep{Norcia3} in Figure~\ref{fig:echoes-beats}. By comparing (a) with the lower part of (b), we see several dominant peaks with several small ripples in both cases, suggesting that they might be caused by the same dynamics. Furthermore, by comparing the upper and lower part of Figure~\ref{fig:echoes-beats}(b), we observe that the superradiant echoes with wild separation are actually evolved from the superradiant beats when the number of sub-ensembles is increased. This observation motivates us to establish the connection as shown in Fig. 4 of the main text.  

\section{First-order Mean-field Equations \label{sec:1stMean}}

In this section, we present the first-order mean-field equations, which are used in the main text. In a frame rotating with the microwave driving field, the intra-cavity field amplitude $\langle \hat{a}\rangle$ satisfies the equation 
\begin{equation}
\partial_{t}\langle \hat{a}\rangle=-i(\tilde{\omega}_{c}-\omega_{d})\langle \hat{a}\rangle-i \sqrt{\kappa_{1}}\Omega(t)-i\sum_{\alpha=1}^{N}g_{\alpha}N_{\alpha}\langle \hat{\sigma}_{\alpha,1}^{12}\rangle . \label{eq:a-1}
\end{equation}
Here, the complex frequency is defined as $\tilde{\omega}_c=\omega_c-i\kappa/2$. In Eq.~\eqref{eq:a-1}, $\langle \hat{\sigma}_{\alpha,1}^{12} \rangle$ describes the coherence of the first spin of the $\alpha$-th sub-ensemble, where the superscripts $1,2$ indicate the lower and higher spin levels. Then, the spin coherence satisfies the equation
\begin{align}
\partial_{t}\langle\hat{\sigma}_{\alpha,1}^{12}\rangle=-i(\tilde\omega_{\alpha}-\omega_{d})\langle\hat{\sigma}_{\alpha,1}^{12}\rangle-ig_{\alpha}\langle \hat{a}\rangle(1-2\langle\hat{\sigma}_{\alpha,1}^{22}\rangle).  \label{eq:sigma12-1}
\end{align}
Here, the complex frequency is defined as $\tilde\omega_{\alpha}=\omega_{\alpha}-i(\gamma_\alpha+\eta_\alpha/2+\chi_{\alpha})$. Eq.~\eqref{eq:sigma12-1} depends on the population of the upper spin level $\langle\hat{\sigma}_{\alpha,1}^{22}\rangle$, which obeys the equation
\begin{align}
&\partial_{t}\langle\hat{\sigma}_{\alpha,1}^{22}\rangle=\gamma_\alpha-(2\gamma_\alpha+\eta_\alpha)\langle \hat{\sigma}_{\alpha,1}^{22}\rangle \nonumber \\
& +ig_{\alpha}(\langle \hat{a}^{\dagger}\rangle\langle\hat{\sigma}_{\alpha,1}^{12}\rangle-\langle \hat{a}\rangle\langle\hat{\sigma}_{\alpha,1}^{21}\rangle). \label{eq:sigma22-1}
\end{align}
In this equation, the terms $\langle \hat{a}^{\dagger}\rangle,\langle\hat{\sigma}_{\alpha,1}^{21}\rangle$ are the complex conjugate of $\langle \hat{a}\rangle,\langle\hat{\sigma}_{\alpha,1}^{12}\rangle$. Since the spins within the same sub-ensemble are assumed to have identical properties, all the terms $\langle \hat{\sigma}_{\alpha,k}^{mn} \rangle$ (for $m,n=1,2$) have the same value and thus they can be represented by the terms of the first spin $\langle \hat{\sigma}_{\alpha,1}^{mn} \rangle$. We solve the above equations numerically and present the corresponding codes in Section~\ref{sec:codes}.

\begin{figure*}
\begin{centering}
\includegraphics[scale=1]{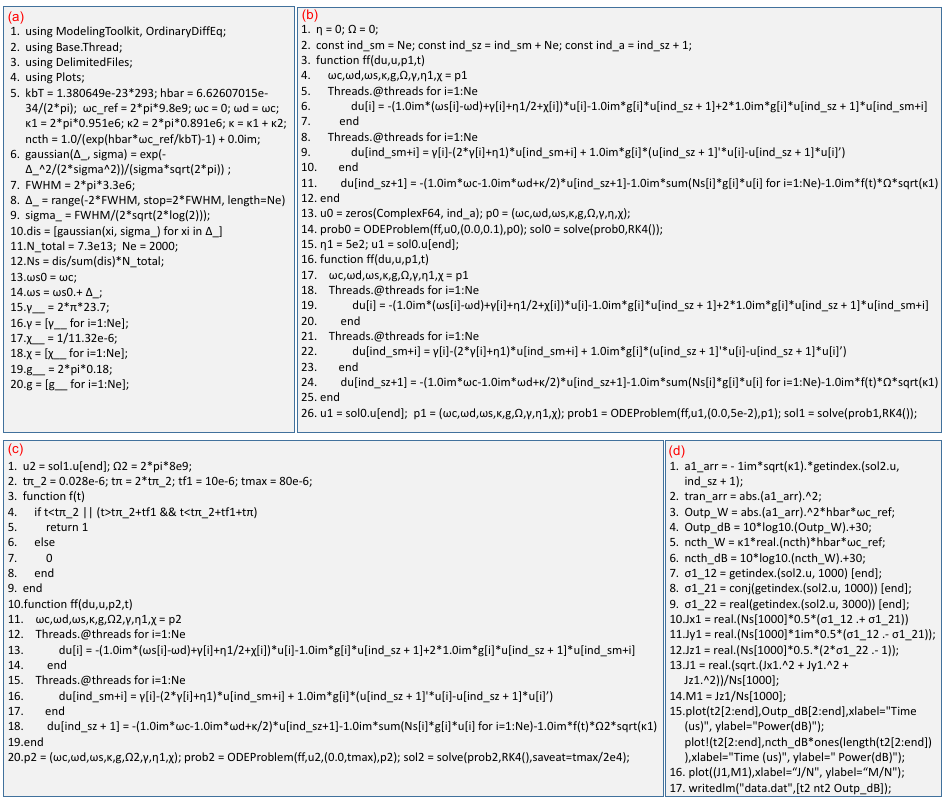} 
\par\end{centering}
\caption{\label{fig:codes}Julia codes to solve the mean-field equations.}
\end{figure*}

\section{\label{sec:codes}Julia Codes to Solve Mean-Field Equations}

In this section, we explain the Julia codes for solving the Eqs. ~\eqref{eq:a-1} to ~\eqref{eq:sigma22-1} with parallel computation as shown in Figure~\ref{fig:codes}. In Figure~\ref{fig:codes}(a),  lines 1 to 4 import the necessary packages. Line 1 employs "ModelingToolkit" to symbolically represent complex systems, while "OrdinaryDiffEq"  is used for solving ordinary differential equation (ODE) problems. Line 2 enables parallel computing, and lines 3 and 4 are used for data storage and plotting, respectively. Line 5 defines the parameters of the microwave resonator along with the thermal equilibrium intra-cavity photon number. Lines 6 to 10 define the Gaussian distribution of the NV center, with line 7 specifically describing the inhomogeneous broadening of the NV centers. Line 11 defines the total number of NV centers and the number of sub-ensembles. Lines 12 to 14 define the number and frequency of NV centers in each sub-ensemble, while lines 15 to 20 specify the spin-lattice relaxation rate, dephasing rate, and the coupling strength between a single NV center and the microwave resonator. 

Firstly, we simulate the equilibrium state achieved by the NV centers and the resonator in the absence of external microwave pumping and laser excitation. In Figure~\ref{fig:codes}(b), line 1 defines the optically induced polarization rate and the microwave pumping rate. Line 2 establishes the indexing scheme for the variables, while lines 3 to 12 define the function that describes the system's time evolution, and is solved using parallel computation methods. Line 13 specifies the initial values and parameters, and line 14 defines the ODE problem (with initial values, time list, parameters), which is solved using the Runge-Kutta (RK4) method. After evolving to an equilibrium state, we define the polarization rate as in line 15 and treat the thermal equilibrium state as the initial state of evolution. Lines 16 to 26 define the same set of equations as in the first stage to bring the system to a cooling steady state.

Then, we begin applying microwave driving. In Figure~\ref{fig:codes}(c), line 1 sets the steady state from the second evolution as the system's initial state and defines the microwave pumping rate. Line 2 specifies the $\pi/2$ and $\pi$  pulses, the free evolution time $\tau$, as well as the total duration of the process. Lines 3 to 9 define the time-dependent function of the microwave pulses, while lines 10 to 20 define and solve the equations. In Figure~\ref{fig:codes}(d), lines 1 and 2 extract the amplitude $\left\langle \hat{a}\right\rangle$ and calculate the intra-resonator photon number. Lines 3 to 6 convert the intra-resonator photon number and the thermal photon number into output power (in dBm). Lines 7 to 9 extract the spin coherence, the spin level population of the 1000th sub-ensemble at the final time. Lines 10 to 12 compute the collective spin vector components, while lines 13 and 14 calculate the average quantum numbers $\bar{J}/N$ and $\bar{M}/N$ of the Dicke state. Lines 15 and 16 plot the dynamics of the output power, thermal noise, and Dicke state, while line 17 saves the data.

In the end of this section, we elaborate the relationship of the Dicke states and the first-order mean-fields. We define firstly the collective operators $\hat{J}_{{x(y)}_\alpha}=\frac{1(i)}{2}\sum_{k=1}^{N_\alpha}\left(\hat{\sigma}_{\alpha,k}^{12}\pm\hat{\sigma}_{\alpha,k}^{21}\right)$, $\hat{J}_{z_\alpha}=\frac{1}{2}\sum_{k=1}^{N_\alpha}\left(2\hat{\sigma}_{\alpha,k}^{22}-1\right)$ for the $\alpha$ sub-ensemble, and then define the Dicke states as the eigen-states of the equations $\left(\sum_{i=x,y,z}\hat{J}_{i_\alpha}^{2}\right)\left|J_\alpha,M_\alpha\right\rangle =J_\alpha\left(J_\alpha+1\right)\left|J_\alpha,M_\alpha\right\rangle $,
$\hat{J}_{z_\alpha}\left|J_\alpha,M_\alpha\right\rangle =M_\alpha\left|J_\alpha,M_\alpha\right\rangle $.  Inspired by the above equations, we introduce the mean values $\bar{J}_\alpha,\bar{M}_\alpha$ through the equations $\bar{J}_\alpha\left(\bar{J}_\alpha+1\right)=\sum_{i=x,y,z}\langle \hat{J}_{i_\alpha}^{2}\rangle $
and $\bar{M}_\alpha=\langle \hat{J}_{z_\alpha}\rangle$~\citep{ZhangY2018,Yuan4}. In the first-order mean field approach, we have $\langle \hat{J}_{i_\alpha}^{2}\rangle \approx \langle \hat{J}_{i_\alpha}\rangle^{2}$. By assuming that the spins are identical in individual sub-ensembles, we can calculate the average values $\bar{J}_\alpha,\bar{M}_\alpha$ with the formulas given in the main text.

\section{\label{sec:eResults}Extra Numerical Results}

In this section, we provide extra numerical results to complement the simulations shown in the main text. 

\subsection{\label{subsec:aDicke} Evolution of Spin Sub-ensembles in Dicke States Space}

\begin{figure}[ht!]
\begin{centering}   
\includegraphics[scale=1]{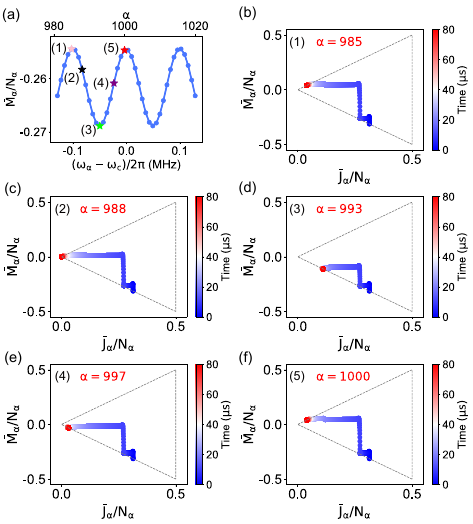}
\par\end{centering}
\caption{\label{fig:dicke} Evolution of the spin sub-ensembles in the Dicke state picture. Panel (a) shows the average excitation number of the Dicke states $\bar{M}_\alpha/N_\alpha$ as a function of the frequency detuning of the spin sub-ensembles to the microwave cavity (lower axis) or the sub-ensemble index $\alpha$ (upper axis) after the second microwave driving pulse. Panels (b)-(f) show the evolution of the spin sub-ensembles in the Dicke states space for five sub-ensembles as marked in the panel (a).}
\end{figure}

In the main text, we observed an excitation grating of spin sub-ensembles after the second microwave driving pulse [Fig. 2(c)], and studied the evolution of the spin sub-ensembles in the inhomogeneous broadening profile [Fig. 2(d)]. To provide a complete view of the dynamics, we consider further several sub-ensembles in a single sub-group [Figure~\ref{fig:dicke}(a)], and present their evolution in the Dicke state space [Figures~\ref{fig:dicke}(b)-(f)]. In principle, all the sub-ensembles behave similarly by following the Hahn echo sequences, except that some sub-ensembles occupy Dicke states with $\bar{M}_\alpha<0$ and some populate the states with $\bar{M}_\alpha>0$ after the second driving pulse. The dephasing causes the sub-ensembles to evolve towards smaller $\bar{J}_\alpha$ horizontally~\citep{ZhangY2018}, the former and latter sub-ensembles evolve to the ground and excited Dicke states along the lower and upper boundary of the Dicke states space, respectively. Since only the former sub-ensembles can transfer the energy to the microwave cavity, the Hahn echo sequences work effectively like a filter, and generate the superradiant echoes with only the excited sub-ensembles.

\subsection{\label{subsec:times} Formation and Evolution of the Phase Grating of Sub-ensembles}

\begin{figure}[ht!]
\begin{centering}   
\includegraphics[scale=1]{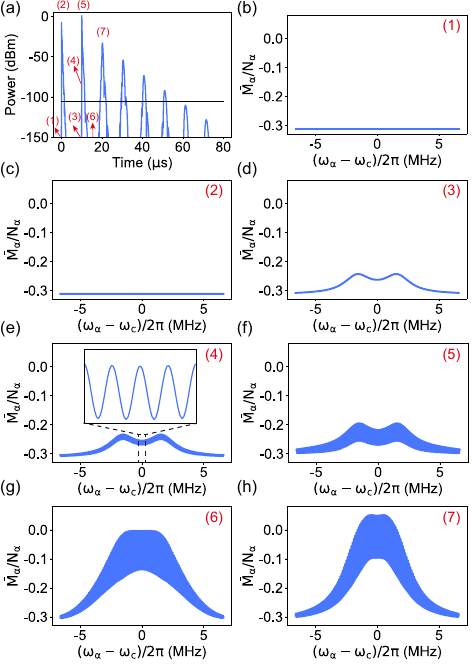}
\par\end{centering}
\caption{\label{fig:times-M/N} Formation and evolution of the phase grating of the spin sub-ensembles. Panel (a) recaps the superradiant echoes as shown in Fig. 2(a), and labels seven special time points. Panels (b)-(h) show the average excitation number of the Dicke states $\bar{M}_\alpha/N_\alpha$ as a function of the frequency detuning of the spin sub-ensemble to the microwave resonator (lower axis) for the seven time points as marked in panel (a).}
\end{figure}

In the previous section, we have selected several sub-ensembles and illustrated their evolution in the Dicke states space. In Figure~\ref{fig:times-M/N}, we consider the evolution of all the spin sub-ensembles at several particular time points as marked in Figure~\ref{fig:times-M/N}(a), and present the average excitation number $\bar{M}_\alpha/ N_\alpha$ (normalized to the number of spins $N_\alpha$) as a function of the frequency detuning of the spin transition to the microwave cavity [Figures~\ref{fig:times-M/N}(b)-(h)]. Initially, the NV spins are optically polarized to the Dicke ground states with $\bar{M}_\alpha =- \bar{J}_\alpha \approx -0.313 N_\alpha$, which leads to the even distribution of $\bar{M}_\alpha$ against frequency detuning [Figure~\ref{fig:times-M/N}(b)]. During the application of the first microwave driving pulse, the microwave cavity is strongly excited, but the spin sub-ensembles are not [Figure~\ref{fig:times-M/N}(c)]. After the first driving pulse, the excited microwave mode starts driving the spin ensembles, and the distribution of $\bar{M}_\alpha$ against frequency detuning shows a double peak structure, reflecting the strong coupling [Figure~\ref{fig:times-M/N}(d)]. During the application of the second microwave driving pulse, the Bloch vectors are rotated along the y-axis [Fig. 2(e)], and the excitation of some sub-ensembles is enhanced and some are suppressed, leading to the excitation grating shown in Figures~\ref{fig:times-M/N}(e)-(f). During the free evolution after the second driving pulse, the spin sub-ensembles evolve to the Dicke states with reduced degree of symmetry $\bar{J}_\alpha$, which reduces the coupling strength with the microwave cavity and the gap between the split peaks of the general feature [Figures~\ref{fig:times-M/N}(g-h)]. 

\subsection{\label{subsec:fre-time} Relationship between Echo Time and Frequency Span of Phase Grating}

\begin{figure}[ht!]
\begin{centering}   
\includegraphics[scale=1]{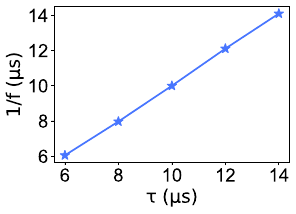}
\par\end{centering}
\caption{\label{fig:fre-time} Linear relationship between the echo time $\tau$ and the inverse frequency span $1/f$ of the phase grating of the spin sub-ensembles.}
\end{figure}

In the main text, we observed a phase grating of the spin sub-ensembles in frequency domain forms after the second microwave driving pulse, and observed that the frequency span $f$ of the phase grating is correlated with the echo time $\tau$. In Figure~\ref{fig:fre-time}, we show in fact that the inverse of the frequency span $1/f$ is identical to the echo time, and thus the phase grating is actually the key to understanding the superradiant echoes effect. 

\subsection{\label{subsec:Jx-times} Spin Sub-ensembles Grating Reflected by Bloch Vector Components}

\begin{figure}[ht!]
\begin{centering}   
\includegraphics[scale=1]{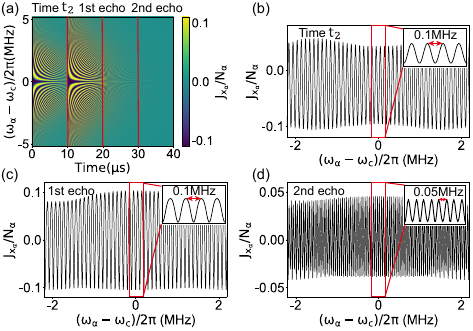}
\par\end{centering}
\caption{\label{fig:Jx-times} Spin sub-ensembles grating. Panel (a) shows the x-component of the Bloch vector $J_{x_\alpha}/N_{\alpha}$ (normalized by the number of spins $N_\alpha$) as a function of the frequency detuning of the spin sub-ensembles to the cavity and the evolution time. Panels (b)-(d) show the cutting lines at the times when the second driving pulse, the first and second echo occur. The insets show the zoom-in in the middle of plots, where the grating period is marked.}
\end{figure}

In Figs. 2(d)-(f), we have analyzed the dynamics of Bloch vectors for spin sub-ensembles when the Hahn echo sequence is applied, and concluded that a phase grating is imprinted among the sub-ensembles at the end of the second driving pulse. In Figure~\ref{fig:Jx-times}, we analyze this grating further by plotting the x component of the Bloch vector $J_{x_\alpha}/N_{\alpha}$ as a function of the frequency detuning of the spin sub-ensembles to the cavity, and the evolution time. Since the y-component of the Bloch vector $J_{x_\alpha}/N_{\alpha}$ shows a similar result, we have not plotted it here. As compared to Fig. 2(a) of Debnath et al.'s paper~\citep{DebnathK}, we have used a yellow-blue gradient color to code the value of $J_{x_\alpha}/N_{\alpha}$ in order to enhance the contrast. While the former figure shows the same color at the echo times, suggesting a re-phasing of all the spin sub-ensembles, our figure shows clearly oscillations as function of the frequency detuning at the echos time, which are demonstrated much more clear in Figures~\ref{fig:Jx-times}(b)-(d). Interestingly, we also find that the grating period is about $0.1$ MHz and $0.05$ MHz for the first and second echo, respectively, and attribute the increase of grating period to the excitation of the spin sub-ensembles on the Dicke ground states by the former echo, as explained in the main text.

\subsection{\label{subsec:Strong} Superradiant Echoes for System with Stronger Coupling}

\begin{figure}[ht!]
\begin{centering}   
\includegraphics[scale=1]{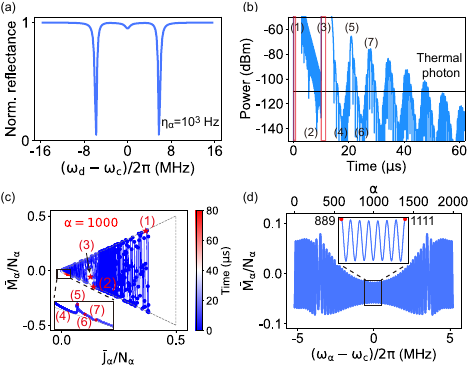}
\par\end{centering}
\caption{\label{fig:Strong} Superradiant echoes in the systems as considered in the experiment~\citep{Putz}, which features a stronger coupling between the NV spins and the microwave cavity. Panel (a) shows the normalized reflectance spectrum with the optical spin cooling rate $\eta_\alpha=10^3$ Hz. Panel (b) displays the superradiant echoes by applying a Hahn echo sequence. Panel (c) shows the evolution of the spin sub-ensemble in the middle of inhomogeneous broadening profile in the Dicke states space, and the numbers (1)-(6) mark the time points as in Fig. 1(b). Panel (d) depicts the average excitation numbers $M_\alpha/N_\alpha$ (normalized by the number of spins $N_\alpha$) as function of the frequency detuning (lower axis) or the index (upper axis) of the spin sub-ensembles after the second pulse. The inset shows the zoom-in of the results for the spin ensembles in the middle of inhomogeneous profile.}
\end{figure}

In the main text, we have mentioned that Putz {\it et al.} have realized for the first time self-stimulated echoes or superradiant echoes with NV centers coupled to a superconducting resonator at cryogenic temperature. In comparison to the experimental system at room temperature~\citep{TomDay} as considered in our study, such a system features much stronger coupling between the NV center spins and the microwave mode. We have applied our theory to simulate such system and obtain the results shown in Figure~\ref{fig:Strong}. Before discussing the results, we comment firstly the parameters. The microwave resonator has a frequency $\omega_c=2\pi\times2.69$ GHz and a damping rate $\kappa=2\pi\times0.8$ MHz as well as a thermal photon number $n_c^{th}=2269$ at $T=293$ K. The microwave driving field has a frequency $\omega_d = 2\pi\times2.69$ GHz and a power $50$ dBm, which leads to a driving strength $\Omega = 2\pi \times 10^{12}$ Hz$^{-1/2}$. The inhomogeneous broadening of NV centers transition frequency is characterized by the Gaussian distribution with a center $\omega_s = 2\pi\times2.69$ GHz, and a linewidth of $2\sqrt{2{\rm ln}2}\sigma = 2\pi\times2.6$ MHz. Each NV center couples to the microwave cavity with a strength $g_\alpha=2\pi\times12$ Hz, and experiences the spin-lattice relaxation, the spin dephasing, and the optically induced spin polarization with rates $\gamma_\alpha = 2\pi\times25$ Hz, $\chi_\alpha = 2\pi\times0.16$ MHz, $\eta_\alpha = 10^3$ Hz, respectively. The durations of $\pi/2$ and $\pi$ pulses are $28$ and $56$ ns, with a free evolution time $\tau$ of $10$ $\mu$s. 

Figure~\ref{fig:Strong}(a) shows the reflection spectrum of a weak microwave probe field, where a splitting of two dips by $2\pi \times 12$ MHz indicates a stronger coupling compared to that in the main text. Figure~\ref{fig:Strong}(b) shows the superradiant echoes when applying the Hahn echo sequence, and each echo consists sharper peaks. Figure~\ref{fig:Strong}(c) shows the evolution in the Dicke states space for an exemplary spin sub-ensemble in the middle of inhomogeneous broadening profile, and this spin ensemble moves up and down multiple times between the lower and upper boundaries when we apply the two microwave drivings and let it progress freely. This dynamics is caused by a faster energy exchange between the NV spins and the microwave photons, and is in strong contrast to the results shown in Fig. 2(b). At the same time, the Bloch vectors rotate multiple circles around some specific axes (not shown), and the dynamics becomes so messy to reveal a clear pattern. In any case, if we look at the average excitation number $\bar{M}_\alpha$ against the frequency detuning of the spin sub-ensembles to the microwave cavity [Figure~\ref{fig:Strong}(d)], we see a butterfly pattern instead of double split peaks, and the zoom-in of a small region in the middle shows a phase grating structure. Thus, we expect that the dynamics leading to the multiple echoes, as illustrated in Fig. 2(c), is still valid here except that this dynamics is strongly modified by the fast energy exchange with the microwave cavity.

\subsection{\label{subsec:couping-cooling} Optically Induced Spin Polarization Modulates Coupling Strength}

\begin{figure}[ht!]
\begin{centering}   
\includegraphics[scale=1]{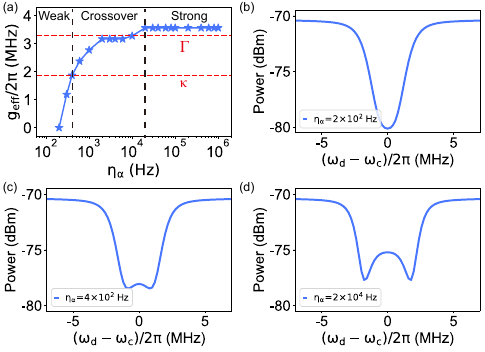}
\par\end{centering}
\caption{\label{fig:couping-cooling} Weak to strong coupling transition. Panel (a) shows the effective coupling strength $g_{eff}$ (blue stars) as a function of optically induced spin cooling rate $\eta_{\alpha}$, and the crossings of this strength with the cavity damping rate $\kappa$ and the linewidth of NV spins' inhomogeneous broadening (lower and upper dashed lines), which distinguishes the weak, crossover and strong coupling regimes. The values of $g_{eff}$ are determined as the frequency gap between two split peaks in the reflection spectrum.  Panels (b)-(d) show the typical reflection spectra for the system in the weak, crossover and strong coupling regime (with $\eta_\alpha=2\times10^2, 4\times10^2$, and $2\times10^4$ Hz), respectively.}
\end{figure}

In Fig. 3(d) we show that the power of the superradiant echoes becomes increased, saturated, and suppressed with increasing optically induced spin cooling rate $\eta_\alpha$, and attribute these changes into the systems in the weak, crossover and strong coupling regimes. To support this claim, in Figure~\ref{fig:couping-cooling}, we study the change of the microwave reflection spectrum for increasing $\eta_\alpha$. By calculating the frequency difference between the two peaks, we extract the effective coupling strength $g_{eff}$, and find that it increases firstly and becomes eventually saturated with increasing $\eta_\alpha$ [Figure~\ref{fig:couping-cooling}(a)], which can be attributed to the change of $\bar{J}_\alpha$~\citep{Yuan1,Yuan2}. By comparing this strength with the damping rate of the microwave cavity $\kappa$ and the linewidth of the NV spins inhomogeneous broadening profile $\Gamma$, we can distinguish precisely the boundaries between the weak, crossover and strong coupling regimes.  As examples, we show the typical spectrum for the systems in these regimes in Figures~\ref{fig:couping-cooling}(b)-(d). These identified boundaries coincide with those in Fig. 3(d), and thus the effect seen in that figure is indeed caused by the system in different regimes. The suppression of the multiple echoes in the strong coupling regimes can be attributed to the photon blockade effect, where the formed spin-photon dressed modes prevent the excitation of the spin ensembles by introducing an off-resonance condition to the microwave driving field.